\documentclass{PoS}

\usepackage{amsmath}
\usepackage{csquotes}
\usepackage{float}
\usepackage{graphicx}

\title{SuperIso Relic new extensions for direct and indirect detection}

\ShortTitle{SuperIso Relic new extensions for direct and indirect detection}

\author{\speaker{G. Robbins}$^{\;1,2,3}$, A. Arbey$^{3,4,}$\thanks{Also Institut Universitaire de France, 103 boulevard Saint-Michel, 75005 Paris, France.}~, F. Mahmoudi$^{3,4,\dagger}$\\
      \\
      $^1$ National Institute of Chemical Physics \& Biophysics, R{\"a}vala
      10, 10143 Tallinn, Estonia \vspace*{0.2cm}\\
        $^2$Univ. Lyon, Univ. Lyon 1, ENS de Lyon, CNRS, Centre de Recherche Astrophysique de Lyon UMR5574, F-69230 Saint-Genis-Laval, France\vspace*{0.2cm}\\
         $^3$ Univ Lyon, Univ Lyon 1, CNRS/IN2P3, Institut de Physique
        Nucl\'eaire de Lyon UMR5822, F-69622 Villeurbanne,
        France\vspace*{0.2cm}\\
       $^4$ CERN, Theoretical Physics Department, CH-1211 Geneva 23,
       Switzerland\\
       \\
        E-mails: \email{glenn.robbins@kbfi.ee}, \email{alexandre.arbey@ens-lyon.fr}, \email{nazila@cern.ch} 
        }

\abstract{SuperIso Relic is a public computing program for the calculation of
flavour observables and relic density in supersymmetry (MSSM and NMSSM). We
present new extensions of the code dedicated to the calculation of dark matter
direct and indirect detection constraints from the latest experimental results.
Contrary to most of the existing programs, this new version allows the user to
consider straightforwardly the uncertainties related to nuclear form factors, dark matter density
and velocity, as well as cosmic-ray propagation through the galactic medium. The
user thus finds a direct way to calculate ``conservative'' , ``standard'' or
``stringent'' constraints according to the chosen set of uncertainties. Some
examplified results showing the impact of such uncertainties are also
presented.}

\FullConference{The 39th International Conference on High Energy Physics (ICHEP2018)\\
		4-11 July, 2018\\
		Seoul, Korea}

\begin{document}

\section{Introduction}

\texttt{SuperIso Relic v4}\footnote{Publicly available on \href{http://superiso.in2p3.fr/relic}{http://superiso.in2p3.fr/relic}} \cite{Arbey:2018msw} is a mixed C / Fortran public program for the calculation of flavour and dark matter observables in the MSSM and NMSSM. It is an extension of the program \texttt{SuperIso}
\cite{Mahmoudi:2007vz} devoted to the calculation of flavour observables that also includes the calculation of dark matter (DM) relic density in standard and modified cosmological scenarios \cite{Arbey:2009gu}. 

Numerous new features were implemented since the previous release, including the possibility of multiprocessor calculation of the relic density, and new cosmological models whose impact on the relic density were studied in detail in \cite{Arbey:2018uho}. In the following, we present the new extensions for the calculation of direct and indirect detection observables of neutralino ($\chi$) dark matter in the (N)MSSM. The new features include
the constraints from recent direct detection experiments (XENON1T \cite{Aprile:2017iyp}, PICO60 \cite{Amole:2017dex}, PANDAX-2 \cite{Cui:2017nnn}), Fermi-LAT dwarf spheroidal galaxy constraints \cite{Fermi-LAT:2016uux} and limits from AMS-02 antiproton data \cite{Aguilar:2016kjl}.
Some of these features are already available in other dark matter public codes such as \texttt{micrOMEGAs}
\cite{Belanger:2004yn}, \texttt{DarkSUSY} \cite{Gondolo:2004sc}, \texttt{MadDM} \cite{Ambrogi:2018jqj} and \texttt{DarkBit} \cite{Workgroup:2017lvb,Athron:2017ard}. In \texttt{SuperIso Relic}, to encourage the users to take into account the underlying astrophysical and nuclear uncertainties, three predefined sets of parameter choices are provided in order to straightforwardly calculate the \enquote{conservative}, \enquote{standard} and \enquote{stringent} constraints for every experiment, as described in \cite{Arbey:2017eos}.

\section{Direct detection}
The implementation of the Spin-Independent (SI) and Spin-Dependent (SD) $\chi$-nucleon scattering amplitudes in the (N)MSSM is based on \cite{Drees:1993bu} and includes QCD and SUSY-QCD corrections \cite{Djouadi:2000ck}.
The likelihoods calculated for each experimental constraints are affected by large nuclear uncertainties which are discussed below.
We refer the reader to \cite{Arbey:2017eos} for a study of the astrophysical uncertainties.

\subsection{Nucleon form factor uncertainties}
\begin{figure}[t]
	\begin{minipage}[t]{0.45\linewidth}
		\includegraphics[width=\linewidth]{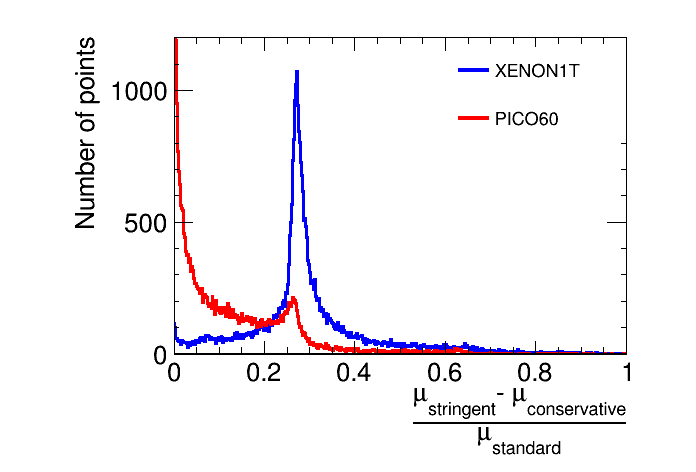}
		\vspace*{-0.8cm}\caption{Relative uncertainties from SI nucleon form factor errors on
			the number of events which would be measured by XENON1T
			and PICO60 for our pMSSM sample of points.}
			\label{fig:nucleonSI}
	\end{minipage}
	\quad\quad\quad
	\begin{minipage}[t]{0.45\linewidth}
		\includegraphics[width=\linewidth]{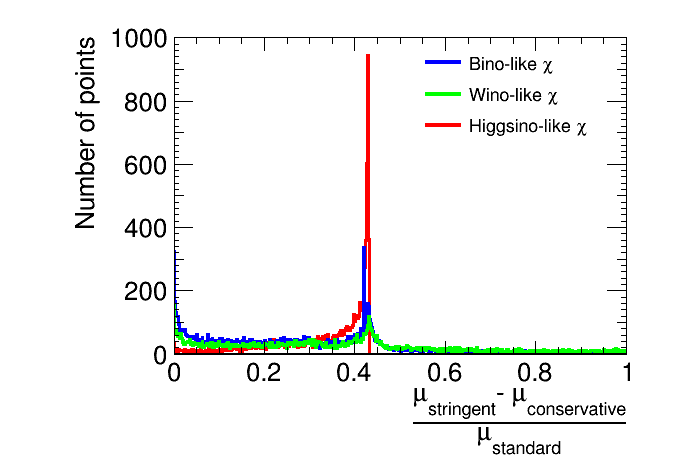}
		\vspace*{-0.8cm}\caption{Relative uncertainties from SD nuclear structure factor errors
			on the number of events which would be measured by PICO60 for our
			pMSSM sample of points.}
	\label{fig:nuclearSD}
	\end{minipage}
\end{figure}

The SI $\chi$-nucleon scattering amplitudes $A^{SI}$ depend on nucleon form factors $f^{p,n}_{i}$ which can be derived from three quantities involving the light and strange quark content of nucleons: $
	\sigma_{l}\equiv \left <N|\bar{u}u+\bar{d}d |N \right>$, \quad $
	\sigma_{s}\equiv \left <N|\bar{s}s |N \right> $, \quad
	$z\equiv  \left( \big \langle N|\overline{u}u | N \big \rangle -\big \langle N|\overline{s}s |N \big \rangle \right)/ \left( \langle N|\overline{d}d |N \big \rangle -\big \langle N|\overline{s}s |N \big \rangle \right)
	$. 
They are calculated with some errors that need to be considered in the scattering amplitude calculations \cite{Ellis:2018dmb}.
In order to illustrate the impact of those uncertainties, we calculated the number of events per unit of time $\mu$ which would be measured by XENON1T and PICO60 for the sample of pMSSM-19 model points described in \cite{Arbey:2017eos} in the \enquote{conservative}, \enquote{standard} and \enquote{stringent} modes. The results are presented in figure \ref{fig:nucleonSI} and show that, although this source of uncertainty can be most of the time disregarded for fluorine experiments such as PICO60, we have a non-negligible typical error of 27\% for xenon experiments such as XENON1T.

\subsection{Nuclear structure factor uncertainties}

Another source of errors comes from the nuclear structure factors appearing in the calculation of the SD $\chi$-nucleus amplitudes, which suffers from significant uncertainties from two-body currents \cite{Klos:2013rwa}.		
To show the importance of this uncertainty in the MSSM, we performed an analysis similar to the one for nucleon form factor uncertainties. For xenon-type experiments, this source of uncertainty can be safely neglected, however, we show in figure \ref{fig:nuclearSD} that it is not the case for fluorine experiments such as PICO60. Indeed, there is a typical relative error of 40\% in the number of expected events in the latter case. This error is particularly relevant for Higgsino-like neutralinos.

\section{Indirect detection}
\subsection{Fermi-LAT dwarf spheroidal galaxies}

\texttt{SuperIso Relic} calculates a delta-log likelihood using the tabulated bin-by-bin likelihoods released by Fermi-LAT Collaboration for each dSph \cite{fermilat}. We consider two sources of uncertainties. First, for some galaxies, the J-factor is not directly deduced from kinematic observations, but predicted from an empirical relation depending on the distance of the galaxy. The uncertainty on the J-factor is, in this case, arbitrary. We use $log_{10}(\Delta J) =0.8$ dex, 0.6 dex and 0.4 dex for the \enquote{conservative}, \enquote{standard} and \enquote{stringent} options respectively. Second, Fermi-LAT collaboration defines \enquote{conservative},
\enquote{nominal}, and \enquote{inclusive} dSph samples, depending on the ambiguity on the kinematics of the galaxies. The differences between the constraints deduced from these three samples can be substantial and are also taken into account.

\subsection{AMS-02 antiprotons}

In order to set constraints from AMS-02 antiproton data, we consider the two-zone diffusion model described in \cite{Boudaud:2014qra}.
The propagation equation of antiprotons through the galactic medium is solved using a semi-analytical
method, which is faster than full numerical approaches.
In \texttt{SuperIso Relic}, two sources of uncertainties are considered. The first one concerns the DM halo density profile, as the core-cusp problem can significantly alter indirect detection constraints. The antiproton spectra after propagation are therefore directly provided for three different density profiles, namely Einasto, Burkert and NFW profiles, to estimate the impact of this source of uncertainty. The second source of uncertainty comes from the propagation parameters describing the diffusion of antiprotons through the galactic medium. The benchmark sets of propagation parameters \textsc{MED} and \textsc{MAX} are considered in \texttt{SuperIso Relic} and as shown in \cite{Arbey:2017eos}, there is one order of magnitude difference between the two extreme limits, given by the \textsc{MED} model and Burkert halo profile for the weakest constraint and the \textsc{MAX} model and Einasto profile for the strongest.
In addition to the pre-defined models, the user can define his/her own astrophysical model by modifying the values of the propagation parameters and by defining any DM density profile with a cylindrical symmetry.

\section{Conclusion}
We presented the new extensions of the public code \texttt{SuperIso Relic} for direct and indirect dark matter detection in the (N)MSSM. In this new version, the focus is on the impact of the astrophysical and nuclear uncertainties. User-friendly routines are provided to calculate \enquote{conservative}, \enquote{standard} and \enquote{stringent} constraints according to the underlying uncertainties and we described here the effect of the astrophysical and nuclear uncertainties.

\end{document}